\documentclass[twocolumn,prl,superscriptaddress]{revtex4}
\usepackage{graphicx}
\usepackage{helvet}
\usepackage{amsmath}
\usepackage{amstext}

\def\eq#1{Eq.~(\ref{#1})}
\def\fig#1{Fig.~\ref{#1}}

\newcommand{\FIG}[3]{
\begin{figure}[t]
\includegraphics[width=\columnwidth]{#2}
\caption{\label{#1}#3}
\end{figure}}

\begin{document}

\title{Stability of adhesion clusters under constant force}
\author{T. Erdmann}
\affiliation{Max-Planck-Institute of Colloids and Interfaces, D-14424 Potsdam, Germany}
\author{U. S. Schwarz}
\affiliation{Max-Planck-Institute of Colloids and Interfaces, D-14424 Potsdam, Germany}
\affiliation{Institute of Theoretical Physics, University of Leipzig, D-04103 Leipzig, Germany}

\begin{abstract}
  We solve the stochastic equations for a cluster of parallel bonds
  with shared constant loading, rebinding and the completely
  dissociated state as an absorbing boundary. In the small force
  regime, cluster lifetime grows only logarithmically with bond number
  for weak rebinding, but exponentially for strong rebinding.
  Therefore rebinding is essential to ensure physiological lifetimes.
  The number of bonds decays exponentially with time for most cases,
  but in the intermediate force regime, a small increase in loading
  can lead to much faster decay. This effect might be used by
  cell-matrix adhesions to induce signaling events through
  cytoskeletal loading.
\end{abstract}

\maketitle

Biological systems have to be able to change quickly in response to
external stimuli. This is one of the reasons why molecular bonds in
biological system are based on non-covalent interactions and have
short lifetimes of the order of seconds. In order to achieve
long-lived assemblies, cells in multicellular organisms adhere to the
extracellular matrix and to each other through clusters of adhesion
molecules. The number of receptors in adhesion contacts can range from
just a few (e.g.\ for tethering of leukocytes to vessel walls or in
the nascent contacts close to the leading edge of a locomoting cell)
to $\sim 10^5$ (e.g.\ in mature cell-matrix contacts).  Most types of
adhesion clusters are coupled to the cytoskeleton and have to function
under mechanical load, which leads to exponentially increased
dissociation rates \cite{c:bell78}.  Rebinding of broken bonds is
often facilitated by the densely packed arrangement of molecular bonds
in the cluster.  However, if the cluster operates under force, it is
likely to be pulled away by some elastic relaxation process in the
moment the last bond has been broken. Therefore the cluster usually
cannot rebind from the completely dissociated state. Recently, the
interest in the role of force at adhesion clusters has strongly
increased also in the biological community, since it has been shown
that force at cell-matrix adhesions correlates with contact size and
intracellular signaling \cite{c:galb98}.

\FIG{fig:cartoon}
{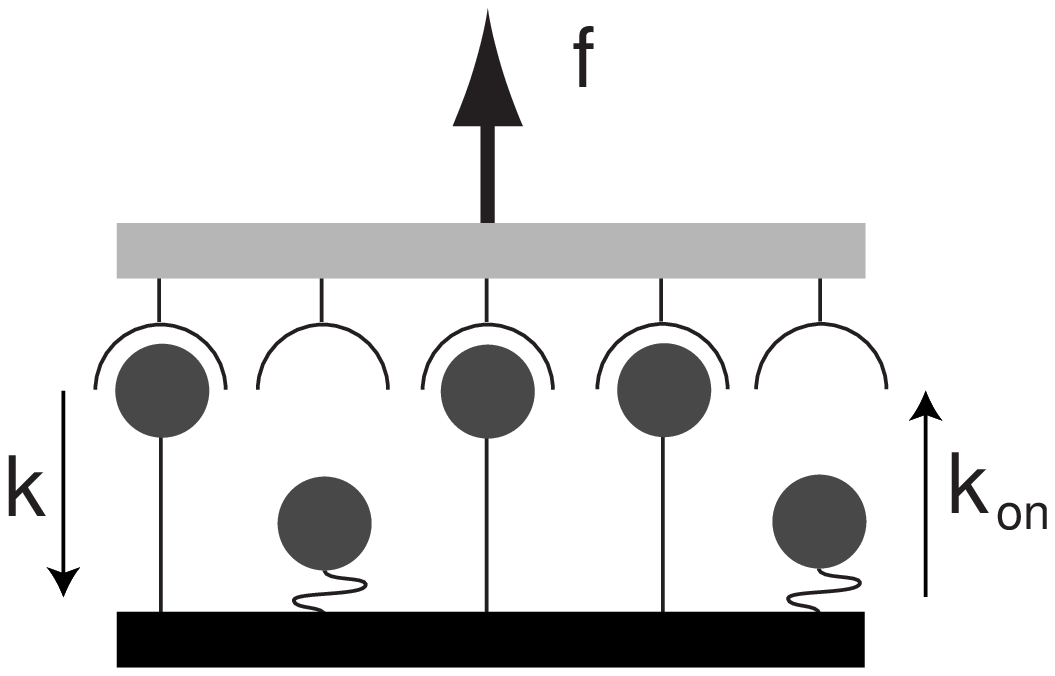} {Schematic representation of an adhesion cluster under
constant force: there are $N_0 = 5$ receptor-ligand pairs, of which $i
= 3$ are closed and equally share the constant dimensionless force
$f$. Single closed bonds rupture with dissociation rate $k = k_0
e^{f/i}$ and single open bonds rebind with force-independent
association rate $k_{on}$. Our model has three parameters: cluster
size $N_0$, dimensionless rebinding rate $\gamma = k_{on}/k_0$ and
dimensionless force $f$.}

Quantitative characterization of adhesion bonds has made tremendous
progress during the last decade, mainly on the level of single
molecules \cite{c:flor94,c:alon95,c:merk99}. Due to the low binding
energies of adhesion bonds, thermal activation is important and
theoretical models are required to interpret experimental data
\cite{c:evan97}. In order to make contact with situations of
biological interest, the quantitative effort now has to be extended to
clusters of adhesion bonds. Clusters also allow to study the effect of
rebinding, which is difficult to address on the level of single
molecules \cite{c:evan01a}. Recently, micropipette techniques have
been used to study cluster dissociation under a linear ramp of force
\cite{c:prec02}, in good agreement with a theoretical analysis by
Seifert \cite{c:seif00}. However, physiological loading of adhesion
clusters is usually more or less constant on the timescale of cluster
lifetime.  The stability of adhesion clusters under constant force has
been first modeled by Bell \cite{c:bell78}, but his treatment was
based on a simplifying deterministic equation for the mean number of
bonds.

In this Letter, we present exact and simulation results for a
stochastic version of the Bell-model. In contrast to the deterministic
model, the stochastic one allows to treat the completely dissociated
state as an absorbing boundary.  Moreover, it includes fluctuation and
non-linear effects which are important for small adhesion
clusters. The main objective of this work is to obtain analytical
results for generic features of adhesion clusters. We present several
new formulae for cluster lifetime as a function of cluster size,
rebinding rate and force, which now can be used for quantitative
analysis of adhesion experiments. Although we do not address the
specific features of cell-matrix adhesions, our model suggests an
appealing mechanism by which cells might induce signaling events
through cytoskeletal loading.

Following Bell \cite{c:bell78}, we consider a cluster with a constant
number $N_0$ of parallel bonds. At any given time, each of the
different bonds can be either open or closed. The constant force $F$
applied to the cluster is assumed to be shared equally between the $i$
closed bonds ($0 \le i \le N_0$). We assume that a single bond under
force $F$ ruptures with the dissociation rate $k = k_0 e^{F / i F_b}$
introduced by Bell \cite{c:bell78}, which can be rationalised by
modelling bond rupture as thermally activated escape over a sharp
transition state barrier \cite{c:evan97}. In this framework, the force
scale $F_b = k_B T / x_b$ is set by thermal energy $k_B T$ and the
distance $x_b$ between the potential minimum and the transition state
barrier along the reaction coordinate of rupture. For typical values
$x_b \sim 1$ nm and $T \sim 300$ K, we find the typical force scale
$F_b \sim 4$ pN. For the single bond association rate $k_{on}$, we
assume that it is independent of force.  Physiological values for both
$k_0$ and $k_{on}$ are expected to be in the $1/s$-range. For the
following, it is useful to introduce dimensionless time $\tau = k_0
t$, dimensionless rebinding rate $\gamma = k_{on} / k_0$ and
dimensionless overall force $f = F / F_b$.  Since bond rupture is a
discrete process, the stochastic dynamics of the bond cluster can be
described by a one-step Master equation \cite{b:kamp92}
\begin{equation} \label{MasterEquation}
\frac{dp_i}{d\tau} = r(i+1) p_{i+1} + g(i-1) p_{i-1} - [ r(i) + g(i) ] p_i
\end{equation}
where $p_i(\tau)$ is the probability that $i$ bonds are closed at time
$\tau$.  The reverse and forward rates between the possible states $i$
are
\begin{equation} \label{Rates}
r(i) = i e^{f / i},\ g(i) = \gamma (N_0 - i)\ .
\end{equation}
Depending on the experimental setup, rebinding from the completely
dissociated state ($i = 0$) might be possible (\textit{reflecting
boundary}) or not (\textit{absorbing boundary}). For $f = 0$ and a
reflecting boundary at $i = 0$, we deal with natural boundaries, that
is given reasonable initial conditions, no special equations are
needed to treat the boundaries. For finite $f$, $r(0) = 0$ is required
to prevent $i$ from becoming negative. An absorbing boundary at $i =
0$ requires $g(0) = 0$.  The mean number of closed bonds $N$ as a
function of time $\tau$ is $N = \langle i \rangle = \sum_{i=0}^{N_0} i
p_i$. From the Master equation \eq{MasterEquation}, one can derive
\begin{equation} \label{FirstMoment}
\frac{d \langle i \rangle}{d\tau} = - \langle r(i) \rangle + \langle g(i) \rangle\ .
\end{equation}
If $r$ and $g$ are both linear functions in $i$, \eq{FirstMoment}
becomes an ordinary differential equation for $N$. This suggests to
study the following deterministic equation
\begin{equation} \label{DeterministicEquation}
\frac{dN}{d\tau} = - N e^{f / N} + \gamma (N_0 - N) 
\end{equation}
as has been done by Bell \cite{c:bell78} for constant loading and by
Seifert \cite{c:seif00} for linear loading. However, for finite
force $f$ solution of \eq{DeterministicEquation} does not give the correct
result for the first moment, since then the rate $r$ defined in
\eq{Rates} is non-linear in $i$ and the average in \eq{FirstMoment}
cannot be taken. More importantly, a differential equation like
\eq{DeterministicEquation} follows from the stochastic 
equations only in the case of natural boundaries. In order to treat
the biologically relevant case of an absorbing boundary, one therefore
has to study the stochastic description \eq{MasterEquation}.

\FIG{fig:pi}
{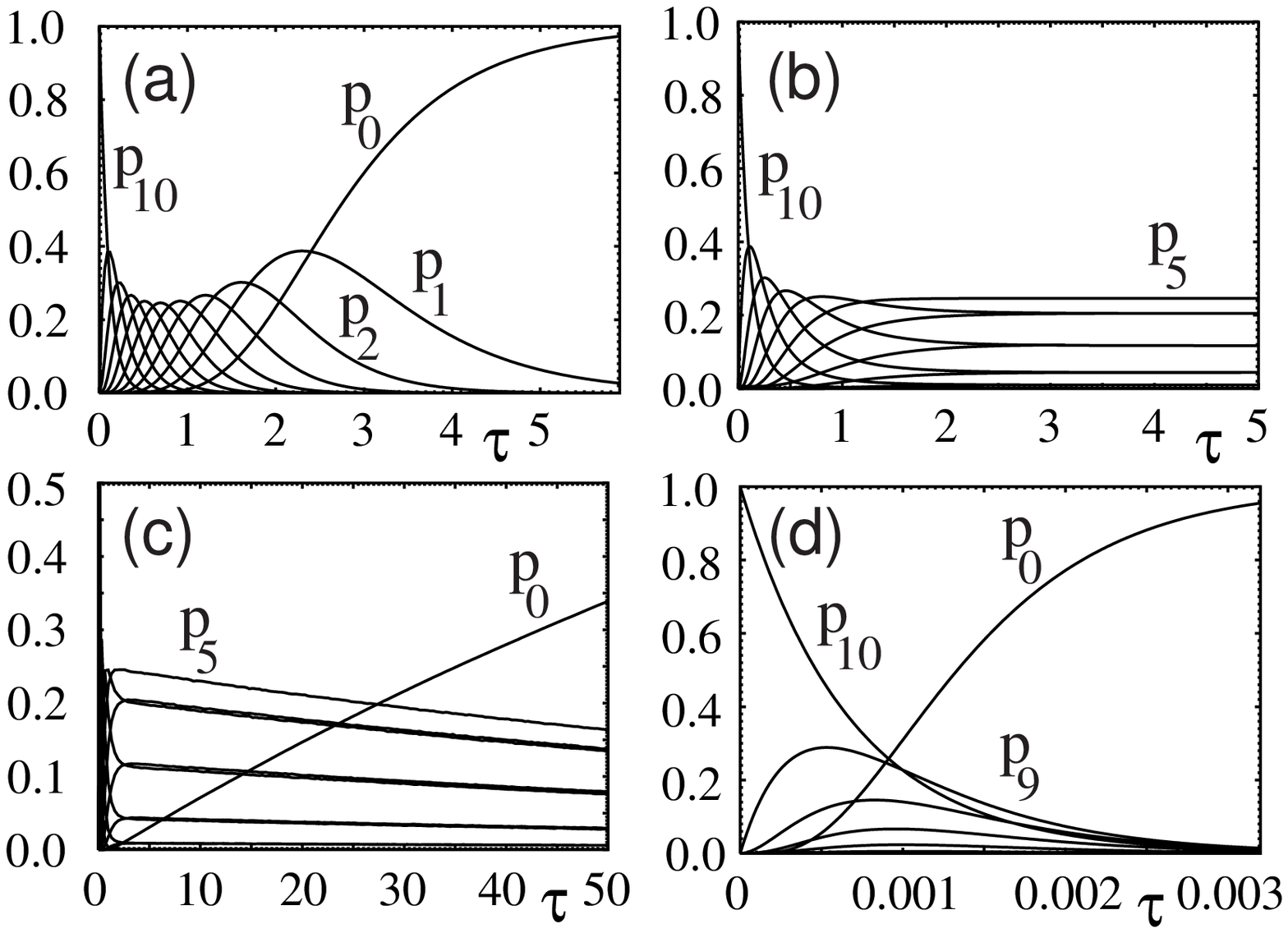} {Solution of the Master equation gives the state
probabilities $p_i$ for $i$ bonds being closed at time $\tau$ ($0 \le
i \le N_0$). Here they are plotted for $N_0 = 10$.  (a) $\gamma = 0$
and $f = 0$.  (b) $\gamma = 1$ and $f = 0$ for a reflecting boundary
at $i = 0$. (c) Same for absorbing boundary. (d) $\gamma = 0$ and $f =
50$. (a) and (b) follow from \eq{pinoforce}, (c) is obtained from
Monte Carlo simulations, and (d) follows from \eq{pinorebinding}.}

A full stochastic solution amounts to finding the set of state
probabilities $p_i(\tau)$ as a function of the three dimensionless
parameters $N_0$, $\gamma$ and $f$.  For force $f = 0$ and a reflecting
boundary, the solution results from the generating function given by
McQuarrie
\cite{r:mcqu63}:
\begin{equation} \label{pinoforce}
p_i(\tau) = \binom{N_0}{i}\frac{\left(\gamma + e^{-(1+\gamma)\tau}\right)^i
\left(1 - e^{-(1+\gamma)\tau}\right)^{N_0 - i}}{(1+\gamma)^{N_0}}\ .
\end{equation}
Here and in the following we use the initial condition $N(0) = N_0$.
If also rebinding $\gamma = 0$, then we deal with the
simple case of independently decaying bonds. \fig{fig:pi}a shows that
in this case, a cluster with $10$ bonds decays from $i = 10$ to $0$ by
visiting each of the intermediate states to an appreciable degree. In
order to stabilize the cluster, one has to introduce rebinding. Then
there is fast relaxation to a stable stationary state, as shown in
\fig{fig:pi}b for $\gamma = 1$.  For the biologically relevant case of
an absorbing boundary, a stable stationary state does not exist and
the cluster will always dissociate on the long run. In this case, one
has to solve the first passage problem of reaching the state $i = 0$
for the first time. This can be done semi-analytically by using
Laplace transforms, where the last backtransform has to be done
numerically.  Alternatively, one can solve the Master equation
numerically by the Monte Carlo method (most efficiently with the
Gillespie algorithm \cite{c:gill77}), as we always do in the general
case, when both rebinding $\gamma$ and force $f$ are
finite. \fig{fig:pi}c shows that in this case, the plateaus from
\fig{fig:pi}b tilt downward, while $p_0$ increases steadily with time
$\tau$. Stability further decreases if force $f$ is turned on. For
very large force, rebinding (including the boundary type at $i
= 0$) becomes irrelevant, because the reverse rate $r$ dominates the
forward rate $g$. Using a recursive scheme to construct $p_i$ from
$p_{i-1}$, for $\gamma = 0$ we find
\begin{equation} \label{pinorebinding}
p_i(\tau) = \left( \prod_{j=i+1}^{N_0} r(j) \right) \sum_{j=i}^{N_0}
 \left\{ \prod_{\substack{k = i\\ k \ne j}}^{N_0} 
\frac{e^{- r(j) \tau}}{r(k) - r(j)} \right\}\ .
\end{equation}
The probability for cluster dissociation at time $\tau$ is 
$p_1(\tau) r(1)$. Setting $i = 1$ in \eq{pinorebinding} and using
\eq{Rates}, one obtains a formula which has been given before in
Ref.~\cite{c:tees01}. \fig{fig:pi}d shows, for the case $f = 50$, that
now the cluster decays very rapidly, with only few of the intermediate
states being visited to an appreciable degree.

\FIG{fig:N}
{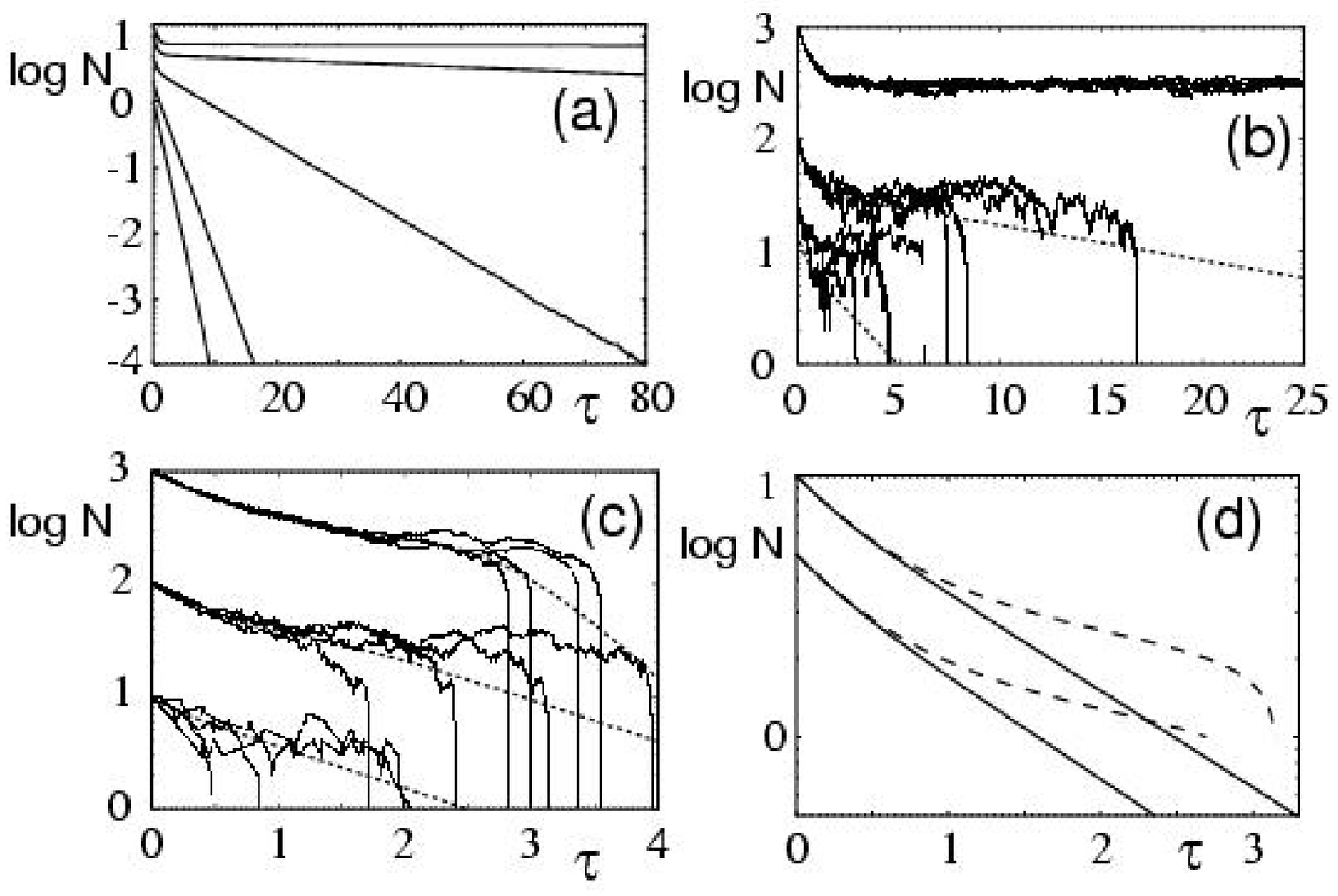} {(a) Simulation results for the mean number of closed bonds
$N$ at time $\tau$ for $f = 0$, $\gamma = 1$ and $N_0 = 1, 2, 5, 10$ and $15$
(lower to upper lines). (b) Four typical simulation trajectories for
each of the cases $N_0 = 10, 100$ and $1000$ for $\gamma = 1$ and
$f/N_0 = 0.25$. Dotted lines are $N(\tau)$.  (c) Same for $f/N_0 =
0.3$. (d) Comparision of stochastic (solid) and deterministic (dashed)
results for $N(\tau)$ for $N_0= 5$ and $10$ for $\gamma = 1$ and
$f/N_0 = 0.3$.}

Once the set of state probabilities $p_i(\tau)$ is known, one can
calculate any quantity of interest, in particular the mean number of
closed bonds $N$ as a function of time $\tau$. We find that $N(\tau)$
usually decays exponentially.  This is demonstrated in \fig{fig:N}a
for the case $f = 0$. The origin of the exponential decay can be
understood as follows: first the system equilibrates into a binomial
distribution peaked around $N_{eq} = \gamma N_0 / (1+\gamma)$ as
described by the result for the reflecting boundary, that is
\eq{pinoforce}. The lower tail of this distribution then 'leaks' into
the state $i = 0$ due to the absorbing boundary. The smaller
$N_0$ or $\gamma$, the larger the tail contribution at $i = 1$ and the
faster the systems loses realizations to the absorbing boundary.  The
resulting decay can be approximated by $N(\tau) \approx N_{eq} e^{-a
\tau}$ with $a \approx p_1(\infty)$ from \eq{pinoforce}. For the 
values of $N_0$ used in \fig{fig:N}a, one finds $a \approx
4.6\times10^{-4}, 9.7\times10^{-3}, 0.16$ and $0.5$ for $N_0 = 2, 5,
10$ and $15$.  Numerically we find $a = 2.5\times 10^{-4}, 8.5 \times
10^{-3}, 0.13$ and $0.6$, thus the leakage estimate is rather good.

In order to assess the role of fluctuations, it is instructive to
study single simulation trajectories.  Since we use the Gillespie
algorithm for exact stochastic simulations \cite{c:gill77}, they are
expected to resemble experimental trajectories. \fig{fig:N}b shows
that for large cluster size and small force, typical trajectories
fluctuate around a plateau value close to $N_{eq}$. However, for small
cluster sizes, fluctuations to smaller bond numbers lead to fast loss
of realizations to the absorbing boundary. Final decay is
rather abrupt due to force-accelerated rupture for decreasing
bond numbers. \fig{fig:N}c shows that for sufficiently large
force, also the large clusters decay quickly.  The loss of stability
for any cluster size follows from Bell's stability analysis of the
deterministic equation \eq{DeterministicEquation}, which yields a
critical force $f_c = N_0\ \rm plog (\gamma / e)$
\cite{c:bell78}, where the product logarithm $\rm plog(a)$ is defined
as the solution $x$ of $x e^x = a$. For typical values of $N_0$ and
$\gamma$, $f_c$ belongs to the intermediate force regime, $1 < f <
N_0$.  For $\gamma = 1$, we have $f_c/N_0 = 0.278$. \fig{fig:N}b and c
are below and above the critical force, respectively.  In contrast to
Bell's continuum analysis, our stochastic analysis shows that for
small clusters a small increase in loading can lead to the fast decay
characteristic for the case without rebinding also for forces below
$f_c$. \fig{fig:N}d compares $N(\tau)$ as obtained from simulations to
$N(\tau)$ as obtained from numerical integration of the deterministic
equation \eq{DeterministicEquation}. This shows that stochastic and
deterministic results differ also on the level of the first moment.

The quantity of largest practical interest is cluster lifetime $T$ as
a function of the model parameters $N_0$, $\gamma$ and $f$. In
general, $T$ can be calculated from the adjoint Master equation
\cite{b:kamp92}. For $N_0 = 2$ and $3$, the solutions can also be found 
by directly summing with appropriate weights over all possible
dissociation paths, each of which is a sequence of Poisson
processes. For $N_0 = 2$ we find
\begin{equation} \label{eq:rebindingexact}
T = \frac{1}{2} \left( e^{-f/2} + 2 e^{-f} + \gamma e^{-3 f/2} \right)\ .
\end{equation}
For $N_0 \ge 4$, the direct procedure becomes intractable. However, in
the case of vanishing rebinding ($\gamma = 0$), there is only one
dissocation path and the exact solution is simply $T = \sum_{i =
1}^{N_0} 1 / r(i)$ for all values of $N_0$ \cite{c:tees01}.  In
\fig{fig:T}a we plot $T$ as a function of $f / N_0$ for different
cluster sizes $N_0$. In the small force regime, $f < 1$, $T$ plateaus
at the value $H_{N_0} = \sum_{i=1}^{N_0} 1/i \approx \ln N_0 + 1/(2
N_0) + \Gamma$. Here $H_{N_0}$ are the harmonic numbers and $\Gamma = 0.577$
is Euler's constant. In this regime, $T$ depends only weakly
(logarithmically) on $N_0$ and large cluster sizes are required to
achieve long lifetimes \cite{c:gold96,c:tees01}. In the intermediate
force regime, $1 < f < N_0$, we find $T \approx H_{N_0} - H_{f}
\approx \ln (N_0 / f)$. Here the effective cluster size is reduced to
$N_0/f$, because the cluster dissociates very rapidly for $i < f$. In
the high force regime, $f > N_0$, only the term with $i = N_0$
contributes: if the first bond breaks, all remaining bonds break
within no time. The destabilizing effect of force can be counteracted
by rebinding. In the case of vanishing force ($f = 0$), the solution
can also be found by using Laplace transforms \cite{b:kamp92}.  We
find
\begin{equation} \label{eq:VF_lifetime}
T = \frac{1}{1+\gamma}\left(\sum_{n = 1}^{N_0}\left\{\binom{N_0}{n}
\frac{\gamma^n}{n}\right\} + H_{N_0}\right)\ .
\end{equation}
For $\gamma = 0$, we recover the result $T = H_{N_0}$ from above.  For
$N_0 = 2$, we get the result $T = (3+\gamma)/2$ following from
\eq{eq:rebindingexact}. In general, $T$ scales $\sim \gamma^{N_0-1}$
with rebinding rate. In \fig{fig:T}b we plot $T$ as a function of
$N_0$ for different values of $\gamma$. For $\gamma < 1$, the
logarithmic dependence of $T$ on $N_0$ is valid over a wide range of
cluster size. However, for very large clusters, lifetime starts
growing exponentially with $N_0$. For $\gamma > 1$, this strong
increase of $T$ with $N_0$ is found for any value of $N_0$. Therefore
increasing rebinding is much more effective than increasing cluster
size in achieving cluster stability, and essential to ensure
physiological cluster lifetimes with reasonable numbers of bonds. For
example, in the absence of both force and rebinding and if the
lifetime of each bond was $1$ s ($k_0 = 1$ Hz),
\eq{eq:VF_lifetime} predicts that the astronomical number of $\sim
10^{40000}$ independent bonds would be needed to achieve a cluster
lifetime of one day ($T \sim 10^5$ s). In contrast, for $k_{on} = 1$
Hz ($\gamma = 1$), the same cluster lifetime $T$ is achieved by $N_0 =
20$. If rebinding is ten times slower than unbinding ($\gamma = 0.1$),
cluster lifetime $T$ is down to $7$ s and one needs $N_0 = 150$ bonds
to regain a cluster lifetime of one day.  In this way, knowing cluster
lifetime and two out of the three parameters $N_0$, $\gamma$ and $f$
allows to estimate the unknown one.

\FIG{fig:T}
{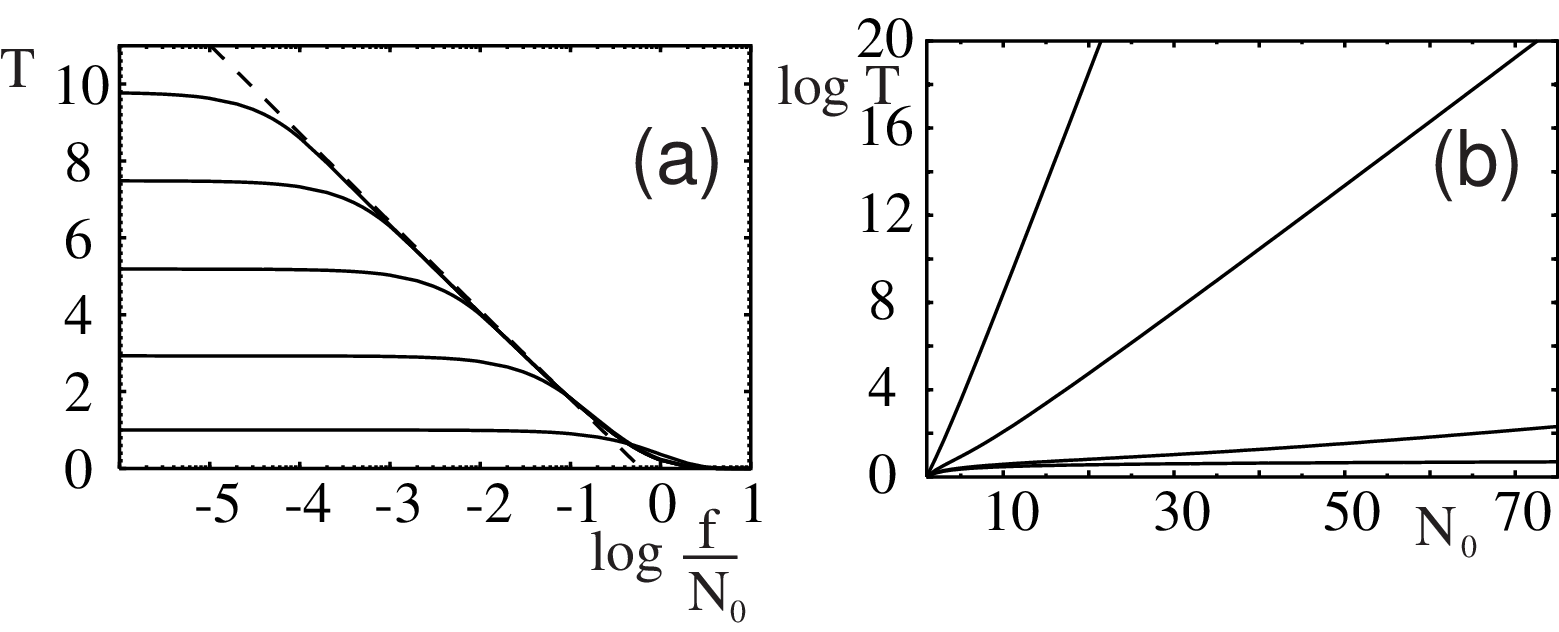} {Analytical results for cluster lifetime $T$.  (a) $T$ as a
function of $f/N_0$ for $\gamma = 0$ and $N_0 = 1, 10, 100, 1000$ and
$10000$ (lower to upper solid lines). The dotted curve is the
approximation $T = \ln\left( 0.61 (N_0/f) \right)$.  (b) $T$ as a
function of cluster size $N_0$ for $f = 0$ and $\gamma = 0.0, 0.1,
1.0$ and $10.0$ (lower to upper lines).}

Our model is also relevant for cell adhesion if initial loading is
much faster than cluster lifetime (otherwise the assumption of
constant force is not valid) and if cluster decay is much faster than
potential reinforcement process (otherwise the assumption of constant
cluster size is not valid). One example which might satisfy these
conditions is L-selectin mediated leukocyte tethering in shear flow
\cite{uss:dwir03a}. Our assumptions do certainly not hold for
cell-matrix adhesions, which have been shown to grow rather than to
decay under the effect of force \cite{c:galb98}. Although the specific
processes at work at cell-matrix processes are not the subject of this
work, our model suggests that the stress constant $\sim 5.5$ nN/$\mu
m^2$ recently measured on elastic substrates for the physiological
loading of cell-matrix contacts through the cell's own contractile
machinery \cite{uss:bala01} might be close to the critical force $f_c
= N_0\ \rm plog (\gamma / e)$, because then small changes in
cytoskeletal loading would result in strongly accelerated cluster
decay. Increased force on a subset of bonds might in turn induce
signaling events (possibly through mechanical opening-up of protein
domains) leading to subsequent recruitment of additional bonds.
Recent single molecule experiments for activated $\alpha_5
\beta_1$-integrin binding to fibronectin gave $k_0 = 0.012$ Hz and
$F_b = 9$ pN \cite{c:li03}.  Setting $F_c = 5.5$ nN and using $N_0 =
10^4$, we predict $\gamma = 0.2$, corresponding to a rebinding rate
$k_{on} = 0.002$ Hz.

This work was supported by the German Science Foundation through the
Emmy Noether Program.


\end{document}